\documentclass[prd,amsmath,amssymb,notitlepage,nofootinbib,onecolumn,10pt]{revtex4-1} 
\usepackage[english]{babel}
\usepackage[latin1]{inputenc}
\usepackage{bbm}
\usepackage{bm}
\usepackage{amsfonts}
\usepackage{graphics,graphicx,calc,epsfig,pstricks}
\newcommand{\be}{\begin{equation}}
\newcommand{\ee}{\end{equation}}
\newcommand{\bea}{\begin{eqnarray}}
\newcommand{\eea}{\end{eqnarray}}

\begin{document}
\title{Spacetime defects and group momentum space}

\author{Michele Arzano}
\email{michele.arzano@roma1.infn.it}
\affiliation{Dipartimento di Fisica and INFN,\\ ``Sapienza" University of Rome,\\ P.le A. Moro 2, 00185 Roma, Italy}

\author{Tomasz Trze\'{s}niewski}
\email{tomasz.trzesniewski@ift.uni.wroc.pl}
\affiliation{Institute for Theoretical Physics,\\ University of Wroc\l{}aw,\\ Pl.\ M.\ Borna 9, 50-204 Wroc\l{}aw, Poland}

\begin{abstract}
\begin{center}
{\bf Abstract}\\
\end{center}
We study massive and massless conical defects in Minkowski and de Sitter spaces in various spacetime dimensions. The energy-momentum of a defect, considered as an (extended) relativistic object, is completely characterized by the holonomy of the connection associated with its spacetime metric. The possible holonomies are given by Lorentz group elements, which are rotations and null rotations for massive and massless defects respectively. In particular, if we fix the direction of propagation of a massless defect in $n\!+\!1$-dimensional Minkowski space, then its space of holonomies is a maximal abelian subgroup of the ${\rm AN}(n-1)$ group, which corresponds to the well known momentum space associated with the $n$-dimensional $\kappa$-Minkowski noncommutative spacetime and $\kappa$-deformed Poincar\'{e} algebra. We also conjecture that massless defects in $n$-dimensional de Sitter space can be analogously characterized by holonomies belonging to the same subgroup. This shows how group-valued momenta related to four-dimensional deformations of relativistic symmetries can arise in the description of motion of spacetime defects.
\end{abstract}


\maketitle
\section{Introduction}
Conical spacetime defects were first introduced by Staruszkiewicz \cite{Staruszkiewicz:1963} as point particles coupled to gravity in 2+1 spacetime dimensions. Several years later Deser, Jackiw and 't Hooft \cite{Deser:1984} generalized this result to configurations of arbitrary many particles. The peculiarity of such systems is that particles are represented by  topological defects, due to the lack of local degrees of freedom for gravity in three spacetime dimensions, and thus strictly speaking they do not interact gravitationally. The same kind of particle solutions were also shown to exist for non-zero cosmological constant, corresponding to 2+1-dimensional de Sitter or anti-de Sitter space \cite{Jackiw:1984}.

In 3+1 dimensions a conical defect can be obtained simply by replacing the pointlike singularity with a singular one-dimensional object. Such linear defects are known under the name of {\it cosmic strings} and first turned out to be possibly generated during a spontaneous gauge symmetry breaking in the early universe \cite{Kibble:1976}. They were then studied in cosmology as a possible source of primordial density fluctuations \cite{Hindmarsh:1995}, where their contribution was subsequently constrained by observations of the cosmic microwave background, see e.g.\! \cite{Bevis:2010}. Cosmic ``superstrings" can be also produced in string inflation models \cite{Sakellariadou:2009}. At a more abstract level, 't~Hooft has recently employed cosmic strings as ingredients of his {\it piecewise flat} model of gravity \cite{Hooft:2008}. In full analogy with a point particle in 2+1 dimensions, the strings he considers are infinitely thin and straight. The generalization of the concept of a codimension two conical defect to an arbitrary number of spacetime dimensions is then completely straightforward: a straight line is replaced by a hyperplane. Such idealized defects are also the focus of our paper.

A somewhat surprising feature of the description of a point particle in 2+1 gravity (with vanishing cosmological constant) is that its extended momentum space is actually a Lie group, ${\rm SO}(2,1)$, which as a manifold has the form of three dimensional anti-de Sitter space \cite{Matschull:1997}. Interestingly, a non-trivial geometry of momentum space is also one of the effects of (quantum) deformations of relativistic symmetries in four spacetime dimensions, which are at the basis of the so-called {\it doubly special relativity} framework \cite{Amelino:2000,Amelino:2002} and of its most recent incarnation: the {\it relative locality} programme \cite{Amelino:2011,Gubitosi:2013,Freidel:2014}. In this context the role of deformed symmetries within 2+1 gravity has been studied in various approaches since the late 90's \cite{Bais:1998,Bais:2002,Freidel:2005,Noui:2006,Glikman:2014,Arzano:2014,
Kowalski:2014}.

One might wonder whether conical defects in higher dimensional (Minkowski) spaces possess properties similar to the three-dimensional case and, in particular, if their motion can be parametrized by some Lie group related to deformations of relativistic symmetries. To this end let us recall that the best studied example of deformed relativistic symmetries in four dimensions is given by the $\kappa$-Poincar\'{e} (Hopf) algebra \cite{Lukierski:1992}, associated with the non-commutative $\kappa$-Minkowski space \cite{Majid:1994}. The momentum space corresponding to such a deformation is given by the Lie group ${\rm AN}(3)$, which is a subgroup of the five-dimensional Lorentz group ${\rm SO}(4,1)$ \cite{Kowalski:2002,Kowalski:2003}. Therefore, it is of interest to investigate whether this group plays any role in the characterization of spacetime defects. We argue here that, at least to a certain extent, this is indeed the case when one considers {\it light-like}, or {\it massless} defects.

A massless conical defect can be obtained by boosting a (time-like) massive defect to the speed of light. In order to achieve a nontrivial limit one usually applies the boost following a prescription of Aichelburg and Sexl, which was first proposed to derive the gravitational field of a photon from the Schwarzschild metric \cite{Aichelburg:1971}. This method was subsequently extended to other singular null objects, in particular massless cosmic strings \cite{Lousto:1991,Barrabes:2002,Meent:2013}. Some of the null sources coexist with impulsive gravitational waves but this is not the case for the straight strings \cite{Meent:2013}. Moreover, since 3+1-dimensional (anti-)de Sitter space can be represented as a hyperboloid embedded in the 4+1-dimensional flat spacetime, the Aichelburg-Sexl boost can also be utilized \cite{Hotta:1993,Podolsky:1997} to obtain massless particle solutions in (anti-)\break de Sitter space from the Schwarzschild (anti-)de Sitter solution (with particles generating impulsive gravitational waves). The same approach was applied to derive null particle solutions in 2+1-dimensional (anti-)de Sitter space \cite{Cai:1999}. Since in 2+1 dimensions there are no propagating degrees of freedom, in this case particles are not accompanied by gravitational waves.

In this work we analyze how momenta of massless defects in $n\!+\!1$-dimensional Minkowski space can be described by null rotations belonging to the ${\rm AN}(n-1)$ group, the momentum space corresponding to the $n$-dimensional noncommutative $\kappa$-Minkowski space. We also discuss the relation between massless defects in Minkowski space and the same defects in de Sitter space of one dimension less. The structure of the paper is as follows. In Section~II we begin with a discussion of massive conical defects generalized to 4+1-dimensional Minkowski space. Static defects are presented first and then we turn to moving defects. In particular, we describe how such defects are characterized by holonomies of the metric connection, which encode their energy and momentum. In Section~III we explain how massless defects in $n + 1$ dimensions are parametrized by elements of the ${\rm AN}(n-1)$ group, especially for a fixed direction of propagation. In Section~IV we discuss both massive and massless defects in de Sitter space, focusing on the 3+1-dimensional case. We conclude with a summary of our analysis.

\section{Conical defects in Minkowski space}
An intuitive picture of a conical spacetime defect is obtained by considering Minkowski space with a wedge removed and the faces of the wedge ``glued together" to form a cone. The gluing is realized by identifying the opposite faces via a rotation by the {\it deficit angle} characterizing the defect. One can generalize this construction to include identifications of the two faces (not necessarily forming a wedge) via a general Poincar\'{e} transformation, thus producing defects known as ``dislocations" and ``disclinations", adopting the terminology used in the classification of defects in solid media (see \cite{Puntigam:1996} for an extended discussion).

As we will discuss below, conical defects are curvature singularities. Thus a given defect carries some mass/energy on its infinitely thin hyperplane. In 2+1-dimensional gravity the Newton's constant $G$ has the dimension of inverse mass\footnote{We work with units $c = 1$.} (it is the inverse Planck mass) and hence provides a natural mass scale for the theory. The quantity $\mu := Gm$ is the dimensionless ``rest energy per point", which is proportional to the deficit angle of a particle at rest $\alpha = 8\pi Gm$. If we generalize this picture to $n + 1$ dimensions, $G$ will have the dimension of length to the power $n - 2$ times inverse mass. Similarly to the three-dimensional case we can then define the dimensionless energy density $\mu := G\rho$, where $\rho$ is the mass per unit of volume of the defect's hyperplane (e.g.\! length of the string in 3+1 dimensions).

The metric of a conical defect can be written in terms of cylindrical coordinates, in which its geometric properties are most transparent. Here we focus on a defect in 4+1 spacetime dimensions, for reasons which will become clear in the following, but an extension to any number of dimensions is straightforward. In analogy with a conical defect in 2+1 \cite{Deser:1984} and 3+1 dimensions \cite{Meent:2013}, in the five dimensional case the metric has the form
\begin{align}\label{eq:2.1}
ds^2 = -dt^2 + dz^2 + dw^2 + dr^2 + \left(1 - 4\mu\right)^2 r^2 d\phi^2\, .
\end{align}
This metric describes an infinite, flat ``cosmic brane" identified with the $zw$-plane. The brane can be thought of as the ``tip" of a conical defect whose deficit angle $\alpha = 8\pi\mu$ is determined by the (dimensionless) {\it rest energy density} of the brane $\mu \in (0,\frac{1}{4})$. The surrounding spacetime is locally isometric to Minkowski space. Indeed, as one can verify by a straightforward calculation, the Riemann curvature vanishes everywhere outside the brane's world-volume, which is the $tzw$-hyperplane. Consequently, the result of the parallel transport of a vector along an arbitrary loop encircling the defect depends only on the conical curvature singularity and the loop's winding number. The parallel transport around a loop $\gamma(\lambda)$ is measured by the {\it holonomy} of the Levi-Civita connection given by
\begin{align}\label{eq:2.2}
(h_\gamma)^\mu_{\ \nu} = {\cal P} \exp\left(-\int_\gamma \Gamma^\mu_{\ \sigma\nu} \frac{d\gamma^\sigma}{d\lambda} d\lambda\right)\,,
\end{align}
where ${\cal P}$ denotes the path ordering and $\Gamma^\mu_{\ \sigma\nu}$ are Christoffel symbols associated with the metric (\ref{eq:2.1}). One may notice that in four space dimensions it is possible to take a loop around a two-dimensional plane due to the fact that closing a path around a given hyperplane requires at least two directions orthogonal to it. The holonomy (\ref{eq:2.2}) of a loop containing the defect is a non-trivial Lorentz transformation, which can be understood as a result of the presence of a delta-peaked curvature on the defect's world-volume \cite{Regge:1961}. Not surprisingly, this Lorentz transformation is also gluing the faces of the defect's wedge \cite{Matschull:1997}.

A generalization of the holonomy (\ref{eq:2.2}) can be used to characterize both position and momentum of a (moving) defect. This is achieved by considering the parallel transport of a coordinate frame instead of just a vector. If the defect is displaced from the frame's origin, the resulting {\it Poincar\'{e} holonomy} in addition to a Lorentz transformation contains also a translation, carrying information about the defect's position \cite{Hooft:2008,Meent:2011}. However, here we will restrict our considerations to Lorentz holonomies (\ref{eq:2.2}), which completely describe the energy-momentum density of a moving defect, as it was first shown for 2+1 dimensions \cite{Matschull:1997}. Let us start by explaining how the rest energy density $\mu$ appearing in the conical metric is encoded in the Lorentz holonomy of a static defect. We do this by deriving the explicit form of the holonomy (\ref{eq:2.2}) corresponding to the metric (\ref{eq:2.1}). The calculation is most easily carried out by picking a simple loop, e.g.\! a circle parametrized by $r = (1 - 4 \mu)^{-1}$, $\phi = 2\pi \lambda$ and $t,z,w = 0$. The result written in terms of Cartesian coordinates $x = r \cos\phi$ and $y = r \sin\phi$ is given by
\begin{align}\label{eq:2.3}
h_R(\alpha) = \left(
\begin{array}{ccccc}
1\! & 0\! & 0\! & 0\! & 0 \\ 
0\! & \cos(\alpha)\! & -\sin(\alpha)\! & 0\! & 0 \\ 
0\! & \sin(\alpha)\! & \cos(\alpha)\! & 0 \! & 0 \\ 
0\! & 0\! & 0\! & 1\! & 0 \\ 
0\! & 0\! & 0\! & 0\! & 1
\end{array}
\right)\,.
\end{align}
Such a holonomy is simply a rotation (an {\it elliptic} Lorentz transformation) by the deficit angle $\alpha = 8\pi\mu$ around the origin in the $xy$-plane.

The metric of a moving defect can be obtained \cite{Meent:2013} by ``boosting" the static metric (\ref{eq:2.1}) in Cartesian coordinates using an ordinary boost in a direction which lies in the $xy$-plane. Indeed, the metric (\ref{eq:2.1}) is obviously invariant under boosts in the directions $z$ and $w$ and therefore the defect can only move in the plane perpendicular to itself, like a point particle in 2+1 dimensions. For simplicity we may take a boost in the $x$ direction
\begin{align}\label{eq:2.4}
t \mapsto t \cosh\chi + x \sinh\chi\,, \nonumber\\ 
x \mapsto x \cosh\chi + t \sinh\chi\,,
\end{align}
where $\chi$ is the rapidity parameter. Introducing lightcone coordinates $u$, $v$ via $t = \frac{v - u}{\sqrt{2}}$, $x = \frac{v + u}{\sqrt{2}}$ we obtain the metric
\begin{align}\label{eq:2.5}
ds^2 = 2 du dv + dy^2 + dz^2 + dw^2 - \left(1 - (1 - 4\mu)^2\right) \frac{(e^\chi (u dy - y du) + e^{-\chi} (v dy - y dv))^2}{e^{2\chi} u^2 + e^{-2\chi} v^2 + 2 (uv + y^2)}\,,
\end{align}
describing a massive brane travelling with velocity $V = \tanh\chi$ in the direction $x$. It can be shown after a simple calculation \cite{Meent:2011} that the deficit angle $\alpha^\prime$ of a moving defect becomes wider than the deficit angle $\alpha$ at rest and that they are related by the formula $\tan(\alpha^\prime/2) = \tan(\alpha/2) \cosh\chi$. Here we observe that in the limit of small $\alpha$ and $\alpha^\prime$ the latter simplifies to
\begin{align}\label{eq:2.5a}
\alpha^\prime = \alpha \cosh\chi\,,
\end{align}
which can be interpreted as the familiar expression for the relativistic energy density. Namely, as we already explained, the rest energy density ${\cal E}_0 = \mu \sim \alpha$, while the factor $\cosh\chi = (1 - V^2)^{-1/2}$ and hence the total energy density ${\cal E}_0 (1 - V^2)^{-1/2} = {\cal E} \sim \alpha^\prime$.

One can also notice that, due to the nontrivial parallel transport of frames, the presence of a conical singularity in the otherwise flat spacetime introduces an ambiguity in the direction of a boost as perceived by different observers. To resolve this issue we have to consistently choose one frame for defining boosts, e.g.\! such one that the defect's wedge lies symmetrically behind the boost's direction. On the other hand, the whole problem can be avoided by directly boosting the holonomy (\ref{eq:2.3}). This is achieved by acting on the rotation $h_R(\alpha)$ by a boost $b(\chi)$ via the conjugation $b^{-1}(\chi) h_R(\alpha) b(\chi)$. The resulting Lorentz holonomy
\begin{align}\label{eq:2.5b}
h_R(\alpha,\chi) = \left(
\begin{array}{ccccc}
1 + 2\sinh^2(\chi) \sin^2(\alpha/2)\! & -\sinh(2\chi) \sin^2(\alpha/2)\! & -\sinh(\chi) \sin(\alpha)\! & 0\! & 0 \\ 
\sinh(2\chi) \sin^2(\alpha/2)\! & 1 - 2\cosh^2(\chi) \sin^2(\alpha/2)\! & -\cosh(\chi) \sin(\alpha)\! & 0\! & 0 \\ 
-\sinh(\chi) \sin(\alpha)\! & \cosh(\chi) \sin(\alpha)\! & \cos(\alpha)\! & 0 \! & 0 \\ 
0\! & 0\! & 0\! & 1\! & 0 \\ 
0\! & 0\! & 0\! & 0\! & 1
\end{array}
\right)
\end{align}
is thus an element of the {\it conjugacy class} of rotations by the angle $\alpha$, which fully characterizes the motion of a defect described at rest by the metric (\ref{eq:2.1}).

\subsection{Massless defects}
So far we have dealt with massive defects. Not surprisingly, it is also possible to consider {\it massless} (lightlike) defects, moving with the speed of light. This can be achieved by deriving the theoretical limiting case of the calculation described above, which consists in performing an ``infinite" boost via the prescription first introduced by Aichelburg and Sexl \cite{Aichelburg:1971} in the case of the Schwarzschild metric. The trick is to keep fixed the quantity $\varrho \equiv 8\pi\mu \cosh\chi$ (which is the laboratory energy density, as can be shown by a simple calculation \cite{Meent:2013}) while taking the limit of the rapidity $\chi \rightarrow \infty$. If we use the well known distributional identity
\begin{align}\label{eq:2.6}
\lim_{b \rightarrow 0} \frac{b}{a^2 + b^2} = \pi \delta(a)\,,
\end{align}
the metric (\ref{eq:2.5}) in the Aichelburg-Sexl limit becomes
\begin{align}\label{eq:2.7}
ds^2 = 2 du dv + dy^2 + dz^2 + dw^2 - \sqrt{2}\varrho |y| \delta(u) du^2
\end{align}
and it describes the geometry of a massless cosmic brane. One might be worried about the distributional nature of this metric. However, solutions to Einstein equations involving distributions have a long history and have been extensively studied (see e.g.\! \cite{Steinbauer:2006}). The metric (\ref{eq:2.7}) corresponds to a $zw$-plane moving along the null direction $v$. Spacetime is still flat outside the defect's world-volume and hence the holonomy of a loop around the defect reflects the presence of a curvature singularity at the hyperplane $u,y = 0$. In the considered coordinates it is convenient to take a square loop with $u,y \in [-1,1]$ and transform the associated holonomy to Cartesian coordinates, which gives
\begin{align}\label{eq:2.8}
h_P(\varrho) = \left(
\begin{array}{ccccc}
1 + \tfrac{\varrho^2}{2}\! & -\tfrac{\varrho^2}{2}\! & -\varrho\! & 0\! & 0 \\ 
\tfrac{\varrho^2}{2}\! & 1 - \tfrac{\varrho^2}{2}\! & -\varrho\! & 0\! & 0 \\ 
-\varrho\! & \varrho\! & 1\! & 0 \! & 0 \\ 
0\! & 0\! & 0\! & 1\! & 0 \\ 
0\! & 0\! & 0\! & 0\! & 1
\end{array}
\right)\,.
\end{align}
This holonomy is a {\it null rotation} by the deficit angle $\varrho$ with respect to the null axis $v$ in the $uy$-plane. Such Lorentz transformations are called {\it parabolic} since their orbits are parabolae on a given lightlike plane, i.e.\! open curves, in contrast to the circular orbits of usual rotations. Null rotations can be expressed by the homogeneous combinations of boosts and usual rotations, as we will discuss in detail in the next Section.

An alternative construction of the metric of a massless defect, free from the ``ambiguity" associated with performing a boost in the conical spacetime, was given in \cite{Meent:2013}. As mentioned above, one can perform a boost $b(\chi)$ of the holonomy describing a static defect (\ref{eq:2.3}) via the conjugation $b^{-1}(\chi) h_R(\alpha) b(\chi)$. If we take the Aichelburg-Sexl limit, the resulting Lorentz group element will be a null rotation (\ref{eq:2.8}). The metric leading to such a holonomy can be then reconstructed following the example of a massive defect (\ref{eq:2.1}). The geometries will be similar but now the deficit angle has to be cut out from a null plane instead of a spacelike one. To this end we need cylindrical-like lightcone coordinates, which can be derived starting from usual lightcone coordinates and making a transformation to $q = v + \tfrac{1}{2} y^2 u^{-1}$, $\varphi = y u^{-1}$. However, the ``angular" coordinate $\varphi$ has the infinite range and therefore the cut can not be introduced by a simple rescaling of it, like it is done for $\phi$ in (\ref{eq:2.1}). One rescales $\varphi$ only in the region $u > 0$, obtaining the metric of the form
\begin{align}\label{eq:2.8a}
ds^2 = 2 du dq + dz^2 + dw^2 + \left(1 - f(\varphi) \Theta(u)\right)^2 u^2 d\varphi^2\,, \qquad \int_{-\infty}^{\infty} f(\varphi) d\varphi = \sqrt{2} \varrho\,,
\end{align}
where a smooth function $f(\varphi) < 1$ with the compact support is the analogue of the rescaling factor $4\mu$ in (\ref{eq:2.1}). As one can verify, this metric has the same holonomy as (\ref{eq:2.7}) and thus they are equivalent. Furthermore, (\ref{eq:2.8a}) has the advantage of being smooth away from the $u = 0$ hyperplane and it shows that a massless conical defect is not accompanied by gravitational waves, see \cite{Meent:2013} for details.

In the next Section we will focus on the group theoretic structure needed to characterize the motion of a massless defect. This will lead us to a suggestive connection with the momentum space of certain widely studied models of {\it deformed relativistic symmetries}.

\section{Momentum space of massless defects and the ${\rm AN}(n)$ Lie group}
As discussed in the previous Section, the motion of a conical defect in Minkowski space (with a given number of dimensions) is completely characterized by the Lorentz holonomy associated with a loop encircling the defect. Thus the space of all possible holonomies can be thought of as energy-momentum space of the defect. In order to describe this momentum space in more detail in the case of massless defects we will now parametrize such a defect using spacetime vectors, as it is customarily done for a moving point particle. To get an intuitive picture we start from familiar defects in 3+1 dimensions and then generalize the discussion to higher dimensional cases and in particular to 4+1 dimensions, which we are especially interested in. 

Let us first consider a massless cosmic string in four dimensional Minkowski space that is oriented along the spacelike direction given by the vector ${\bf d} = (0,0,0,1)$ and moving in the null direction ${\bf n} = (1,1,0,0)$ (in Cartesian coordinates $t,x,y,z$). Its motion is completely captured by the holonomy (cf.\! (\ref{eq:2.8}))
\begin{align}\label{eq:3.1x}
h_P(\varrho) = \left(
\begin{array}{cccc}
1 + \tfrac{\varrho^2}{2}\! & -\tfrac{\varrho^2}{2}\! & -\varrho\! & 0 \\ 
\tfrac{\varrho^2}{2}\! & 1 - \tfrac{\varrho^2}{2}\! & -\varrho\! & 0 \\ 
-\varrho\! & \varrho\! & 1\! & 0 \\ 
0\! & 0\! & 0\! & 1 
\end{array}
\right)\,.
\end{align}
The Lorentz group element (\ref{eq:3.1x}) is a null rotation by the angle $\varrho$ with respect to the null axis ${\bf n}$, in the plane spanned by the lightlike vector $(1,-1,0,0)$ and spacelike vector $(0,0,1,0)$. In general, a massless string in 3+1 dimensions can be completely characterized by 4 parameters \cite{Meent:2011}. Indeed, one needs to specify a parabolic angle $\varrho$, carrying the defect's energy density, a lightlike vector ${\bf n}$, along which the defect propagates and a spacelike vector ${\bf d}$, which is the direction of the spatial extension of the defect, with the orthogonality condition ${\bf n} \cdot {\bf d} = 0$ (a defect is invariant under boosts acting along ${\bf d}$). The overall scaling of ${\bf n}$ and ${\bf d}$ is irrelevant. In the end one has $2 + 2 - 1 + 1 = 4$ independent coefficients, which parametrize all possible holonomies. Thus the space of holonomies/momenta of a massless string is bigger than momentum space of a massless particle in 3+1 dimensions, which is only three dimensional. The reason is that conical defects in spacetimes with more than three dimensions are extended objects and the holonomies must account for their non-trivial orientation in space. 

The full space of momenta described above can be restricted in two simple ways. Firstly, we may fix the spatial orientation of a string ${\bf d}$. Then, however, the defect effectively behaves like a point particle in 2+1 dimensions, since it can only move in two directions perpendicular to itself. Secondly, we may choose to fix the direction of motion ${\bf n}$. It turns out that this case is more interesting. To be specific let us introduce a complete orthogonal set of null vectors ${\bf n}_{(x)} = (1,1,0,0)$, ${\bf n}_{(y)} = (1,0,1,0)$, ${\bf n}_{(z)} = (1,0,0,1)$, as well as a set of orthonormal spacelike vectors ${\bf d}_{(x)} = (0,1,0,0)$, ${\bf d}_{(y)} = (0,0,1,0)$, ${\bf d}_{(z)} = (0,0,0,1)$. Suppose that we fix the direction of defect's velocity as ${\bf n} = {\bf n}_{(a)}$, $a \in \{x,y,z\}$. Then the spatial orientation of a defect is orthogonal to ${\bf n}_{(a)}$ and is a linear combination of two vectors ${\bf d}_{(j)}$, $j = 1,2$, which can be chosen to be ${\bf d}_{(b)}$, $b \neq a$. For example, for ${\bf n}_{(x)}$ we may take ${\bf d}_{(1)} \equiv {\bf d}_{(z)}$ and ${\bf d}_{(2)} \equiv {\bf d}_{(y)}$. We observe that for a given ${\bf n}_{(a)}$ the space of holonomies of a massless defect is a subgroup of the Lorentz group ${\rm SO}(3,1)$ determined by the generators of null rotations $X^{(a)}_j := J_{bt} + J_{ba}$, with each $j$ corresponding to one of the spatial indices $b \neq a$ (see also (\ref{eq:3.10}) below). The holonomy of a defect characterized by ${\bf n}_{(a)}$, ${\bf d}_{(j)}$ is obtained by exponentiating such a generator to
\begin{align}\label{eq:3.1}
h^{(a)}_j(k_j) = e^{i k_j X^{(a)}_j}\,, \quad a = x,y,z\,, \quad j = 1,2
\end{align}
(no summation in the exponent), while the parabolic angle $\varrho \equiv k_j$ parametrizes the family of holonomies. For example, if the defect moves along ${\bf n}_{(x)}$, we have two generators $X^{(x)}_1 = J_{yt} - J_{yx}$, $X^{(x)}_2 = J_{zt} + J_{zx}$, which can be written in a four dimensional matrix representation as
\begin{align}\label{eq:3.1ax}
X^{(x)}_1 = i\left(
\begin{array}{cccc}
0\! & 0\! & -1\! & 0 \\ 
0\! & 0\! & -1\! & 0 \\ 
-1\! & 1\! & 0\! & 0 \\ 
0\! & 0\! & 0\! & 0 
\end{array}
\right)\,, \qquad  
X^{(x)}_2 = i\left(
\begin{array}{cccc}
0\! & 0\! & 0\! & -1 \\ 
0\! & 0\! & 0\! & -1 \\ 
0\! & 0\! & 0\! & 0 \\ 
-1\! & 1\! & 0\! & 0 
\end{array}
\right)\,.
\end{align}
The holonomy (\ref{eq:3.1x}) in this picture is given by the group element $h^{(x)}_1 = e^{i k_1 X^{(x)}_1}$, where $\varrho \equiv k_1$, as can be verified by a direct calculation. 

To obtain the holonomy of a defect with a fixed ${\bf n}_{(a)}$ but an arbitrary ${\bf d}$ we just take the product
\begin{align}\label{eq:3.1a}
h^{(a)}_1(k_1) h^{(a)}_2(k_2) = e^{i k^j X^{(a)}_j}
\end{align}
and then the (normalized) spacelike vector is a linear combination ${\bf d} = \sqrt{k_1^2 + k_2^2}^{-1} (-k_2 {\bf d}_{(1)} + k_1 {\bf d}_{(2)})$, while the parabolic angle is given by $\varrho \equiv \sqrt{k_1^2 + k_2^2}$, as one can straightforwardly calculate. Therefore, the defect's energy density is naturally expressed in terms of group coordinates $k_1$, $k_2$, which also specify the defect's orientation. This agrees with the counting of degrees of freedom discussed above since the choice of ${\bf n}$ fixes two parameters and another two remain free. 

Let us now notice that the generators $X^{(a)}_1$ and $X^{(a)}_2$ commute and thus span a {\it maximal abelian subalgebra} of the Lorentz algebra $\mathfrak{so}(3,1)$. To complete the picture we may additionally consider the boost generator $X^{(a)}_0 := J_{0a}$, which in our matrix representation for $a = x$ is given by
\begin{align}\label{eq:3.6}
X^{(x)}_0 = -i\left(
\begin{array}{cccc}
0\! & 1\! & 0\! & 0 \\ 
1\! & 0\! & 0\! & 0 \\ 
0\! & 0\! & 0\! & 0 \\ 
0\! & 0\! & 0\! & 0 
\end{array}
\right)\,. 
\end{align}
(The representations of generators for other choices of ${\bf n}_{(a)}$ are collected in Appendix A.) Group elements generated by a given $X^{(a)}_0$, namely
\begin{align}\label{eq:3.1b}
g^{(a)}_0 = e^{i k_0 X^{(a)}_0}\,, \quad a = x,y,z\,,
\end{align}
form a one-dimensional subgroup of hyperbolic Lorentz transformations in the direction of the spatial component of ${\bf n}_{(a)}$, which can be seen as complementary with respect to null rotations generated by $X^{(a)}_1$, $X^{(a)}_2$. Acting on a holonomy $h^{(a)}_j (k_j)$ via the conjugation
\begin{align}\label{eq:3.5}
g^{(a)}_0 h^{(a)}_j (k_j) (g^{(a)}_0)^{-1} = h^{(a)}_j (e^{k_0} k_j)
\end{align}
one rescales the parabolic angle $\varrho \equiv k_j$ by the factor of $e^{k_0}$, while preserving the orientation of the defect. Thus the effect of a boost is to red- or blueshift the defect's energy density. We should stress that the boost, analogously to the case of massive defects, has to act via conjugation since such an action preserves the trace of a Lorentz group element and consequently the property that the holonomy belongs to the parabolic conjugacy class. 

Finally, we notice that the generator $X^{(a)}_0$ does not commute with $X^{(a)}_1$ and $X^{(a)}_2$, and the respective commutators are given by 
\begin{align}\label{eq:3.5a}
[X^{(a)}_0, X^{(a)}_j] = -i X^{(a)}_j\,, \qquad j = 1,2\,.
\end{align}
The Lie algebra generated by $X^{(a)}_0$, $X^{(a)}_1$, $X^{(a)}_2$ (with a trivial commutator between $X^{(a)}_1$ and $X^{(a)}_2$) is usually denoted by $\mathfrak{an}(2)$ and called the (three-dimensional) {\it abelian nilpotent algebra}, since it possesses an abelian subalgebra spanned by the generators $X^{(a)}_1$, $X^{(a)}_2$, which are also nilpotent, e.g.\! $(X^{(a)}_j)^3 = 0$ in a four dimensional matrix representation. 

The generalization of the above picture to any number of spacetime dimensions is conceptually straightforward. In $n + 1$ dimensions we have $n$ independent null directions, corresponding to $n$ families of generators of null rotations and complementary boosts, labelled by $a = 1,\ldots,n$,
\begin{align}\label{eq:3.10}
X^{(a)}_0 = J_{0a}\,, \qquad X^{(a)}_j = J_{j0} + J_{ja}\,, \quad j \neq a = 1,\ldots,n
\end{align}
where $J_{\mu\nu}$ are generators of the Lorentz group ${\rm SO}(n,1)$. More precisely, $J_{0a}$ generates boosts in the $x_a$-direction, $J_{j0}$ boosts in one of the directions $x_j$ orthogonal to $x_a$ and $J_{ja}$ rotations in the plane spanned by $x_a$ and $x_j$. Similarly to the generators appearing in (\ref{eq:3.5a}), the families (\ref{eq:3.10}) are actually representations of the generators $X_0$, $X_j$ (with nilpotent generators satisfying $(X_j)^3 = 0$ in a $(n+1) \times (n+1)$ matrix representation) of the $\mathfrak{an}(n-1)$ algebra, a subalgebra of the Lorentz algebra $\mathfrak{so}(n,1)$. 

Quite interestingly, the Lie algebra $\mathfrak{an}(n-1)$ is very popular in the noncommutative geometry community, where it is known as the $\kappa$-Minkowski spacetime (first introduced in the four dimensional version in \cite{Majid:1994}) and the generators $X^{(a)}_0$ and $X^{(a)}_j$, after the appropriate rescaling by a dimensionful constant, are identified with time and space coordinates, respectively. Furthermore, the group ${\rm AN}(n-1)$ generated by $\mathfrak{an}(n-1)$ can be obtained from the (local) Iwasawa decomposition of the Lorentz group \cite{Vilenkin:1992} and, in the field theoretic models on $\kappa$-Minkowski space, coordinates on this group are {\it momentum variables} determined by a {\it quantum group Fourier transform} \cite{Guedes:2013,Arzano:2014,Freidel:2007} of functions of non-commuting spacetime coordinates. These momentum coordinates transform under the non-linear boosts and correspond to translation generators of a quantum deformation of the Poincar\'{e} algebra known as the $\kappa$-Poincar\'{e} Hopf algebra \cite{Lukierski:1992,Lukierski:1994}. From the geometric point of view, the Lie group manifold ${\rm AN}(n-1)$ is given by half of $n$-dimensional de Sitter space \cite{Kowalski:2002,Kowalski:2003} and thus the momentum space of such a model is curved, which is a characteristic feature of the recently proposed ``relative locality" approach to modelling phenomenological aspects of quantum gravity \cite{Amelino:2011,Gubitosi:2013}. 

In more than 3+1 dimensions, to completely characterize a given massless defect it is enough to take a parabolic deficit angle, carrying its energy density, and two vectors that determine a two-form normal to the defect's world-volume (which is of codimension 2) \cite{Meent:2011}. The vectors (one lightlike and one spacelike) are mutually orthogonal, their scaling is irrelevant, and thus in $n + 1$ dimensions we deal in total with $2n - 2$ parameters, which should be encoded in the defect's holonomy. If we now consider the restriction imposed on a defect by fixing its direction of motion, as we did in the 3+1 dimensional case, then the holonomy will have $n - 1$ parameters less and hence only $n - 1$ parameters will remain. Therefore, momentum space of a restricted defect is always given by the maximal abelian subgroup of ${\rm AN}(n-1)$, formed by null rotations. 

The most interesting is the 4+1-dimensional case, corresponding to (a subgroup of) the ${\rm AN}(3)$ group, which is the momentum space associated with the $\kappa$-Poincar\'{e} algebra in 3+1 spacetime dimensions. The holonomy of an arbitrary massless cosmic brane with the fixed velocity vector ${\bf n}$ (we drop here the superscript $(a)$) can be written as
\begin{align}\label{eq:3.9b}
h(k_1,k_2,k_3) = h_1(k_1) h_2(k_2) h_3(k_3)\,,
\end{align}
where $h_j$, $j = 1,2,3$ are defined analogously to their four dimensional counterparts (\ref{eq:3.1}), (\ref{eq:3.1b}). As can be shown after some calculations, the energy density of a brane characterized by this holonomy is given by $\varrho = \sqrt{k_1^2 + k_2^2 + k_3^2}$. The brane's spatial plane is spanned by two perpendicular vectors, which can be put in the form ${\bf d}_I = n {\bf d}_{(1)} + n^\prime {\bf d}_{(2)} + {\bf d}_{(3)}$, ${\bf d}_{II} = n^{\prime\prime} {\bf d}_{(1)} - (1 + n n^{\prime\prime})(n^\prime)^{-1} {\bf d}_{(2)} + {\bf d}_{(3)}$, where the coefficients $n = (k_2 - k_3)/(k_1 - k_2)$, $n^\prime = (k_3 - k_1)/(k_1 - k_2)$, $n^{\prime\prime} = (k_2 (k_2 - k_1) + k_3 (k_3 - k_1))/(k_1 (k_1 - k_3) + k_2 (k_2 - k_3))$ and the vectors ${\bf d}_{(j)}$, $j = 1,2,3$ form an orthonormal set orthogonal to ${\bf n}_{(a)}$. Thus there are three parameters $n,n^\prime,n^{\prime\prime}$ specifying the brane as an extended physical object but one of them is irrelevant since ${\bf d}_I$ and ${\bf d}_{II}$ give a purely geometrical internal construction, while it is enough to know a single spacelike vector that is orthogonal to the defect's world volume. 

In the Section below we extend our discussion to (massless) defects in 3+1-dimensional de Sitter space, with cosmological constant $\Lambda > 0$, showing how their spacetime metric is strictly related to the metric of defects in 4+1-dimensional Minkowski space.

\section{Conical defects in de Sitter space}
We look here at a generalization of the derivation of a massless conical defect as it was done in 2+1-dimensional de Sitter space \cite{Cai:1999}. (In Appendix B we have included a similar discussion for defects in anti-de Sitter space.) We begin with a static massive conical defect in 3+1 dimensions \cite{Ghezelbash:2002,Mello:2009}, analogous to a point particle in 2+1 dimensions \cite{Jackiw:1984}, whose metric in static de Sitter coordinates has the form 
\begin{align}\label{eq:2.9}
ds^2 = -(1 - \lambda r^2) d\tau^2 + (1 - \lambda r^2)^{-1} dr^2 + r^2 \left(d\theta^2 + (1 - 4\mu)^2 \sin^2\theta d\phi^2\right)
\end{align}
(where we denote $\lambda \equiv \Lambda/3$) and describes a cosmic string of rest energy density $\mu$, corresponding to the deficit angle $8\pi\mu$. Let us observe that the above metric can be written in terms of embedding Cartesian coordinates of the 4+1-dimensional Minkowski space 
\begin{align}\label{eq:2.11}
t = \sqrt{\lambda^{-1} - r^2} \sinh(\sqrt{\lambda}\tau)\,, \qquad w = \pm\sqrt{\lambda^{-1} - r^2} \cosh(\sqrt{\lambda}\tau)\,, \nonumber\\ 
x = r \cos\phi \sin\theta\,, \qquad y = r \sin\phi \sin\theta\,, \qquad z = r \cos\theta\,,
\end{align}
subject to the $(4,1)$-hyperboloid condition $\lambda^{-1} = -t^2 + x^2 + y^2 + z^2 + w^2$. The transformed (\ref{eq:2.9}) turns out to have the identical form to the metric (\ref{eq:2.1}) of a defect in 4+1-dimensional Minkowski space. There are only two obvious subtleties in the de Sitter case: spacetime coordinates are constrained by the hyperboloid condition and the interpretation of $\mu$ has to be adjusted according to the number of dimensions, since it now represents the {\it linear} energy density rather than energy per surface. A consequence of this situation is that we could indirectly apply the derivation of lightlike defects from Section~II to the de Sitter case. Nevertheless, it may be more illuminating to stick to the approach of \cite{Cai:1999} and see whether we will obtain in this way the same metric as (\ref{eq:2.7}). 

Our goal is to derive a massless defect. Therefore, for convenience we first rescale the radial coordinate $r$ to $r/(1 - 4\mu)$ and expand the metric to the first order in $\mu$, which gives
\begin{align}\label{eq:2.10}
ds^2 \approx ds_{dS}^2 + 8\mu \left(\lambda r^2 d\tau^2 + (1 - \lambda r^2)^{-2} dr^2 +  r^2 d\theta^2\right)\,,
\end{align}
where $ds_{dS}^2$ is the pure de Sitter metric $ds_{dS}^2 = -(1 - \lambda r^2) d\tau^2 + (1 - \lambda r^2)^{-1} dr^2 + r^2 \left(d\theta^2 + \sin^2\theta d\phi^2\right)$. We subsequently transform (\ref{eq:2.10}) to embedding coordinates (\ref{eq:2.11}) and perform a boost $t \mapsto t \cosh\chi + x \sinh\chi$, $x \mapsto x \cosh\chi + t \sinh\chi$, with the rapidity parameter $\chi$, obtaining
\begin{align}\label{eq:2.12}
ds^2 &= ds_{dS}^2 + \frac{8\mu}{(T^2 - w^2)^2} \left((\lambda^{-1} + T^2 - w^2) (w dT - T dw)^2 + \frac{\lambda^{-2} (T dT - w dw)^2}{\lambda^{-1} + T^2 - w^2}\right) \nonumber\\ 
&+ 8\mu \frac{(z (-T dT + z dz + w dw) + (\lambda^{-1} + T^2 - z^2 - w^2) dz)^2}{(\lambda^{-1} + T^2 - w^2)(\lambda^{-1} + T^2 - z^2 - w^2)}\,,
\end{align}
where $T \equiv t \cosh\chi - x \sinh\chi$ (and $ds_{dS}^2$ is now given in embedding coordinates). The metric $ds_{dS}^2$ is preserved by the boost since the latter belongs to the isometry group of de Sitter space. To proceed further we introduce lightcone coordinates $u = (x - t)/\sqrt{2}$, $v = (x + t)/\sqrt{2}$ and employ the Aichelburg-Sexl boost prescription, taking $\chi \rightarrow \infty$ and keeping the laboratory energy density $\varrho := 8\pi\mu \cosh\chi$ constant. Using the distributional identity \cite{Hotta:1993}
\begin{align}\label{eq:2.13}
\lim_{\chi \rightarrow \infty} f(T^2) \cosh\chi = \frac{1}{\sqrt{2}} \delta(u) \int_{-\infty}^{+\infty} f(T^2) dT\,,
\end{align}
we finally obtain
\begin{align}\label{eq:2.14}
ds^2 = 2 du dv + dy^2 + dz^2 + dw^2 - \sqrt{2}\varrho |y| \delta(u) du^2\,.
\end{align}
The metric (\ref{eq:2.14}) corresponds to a conical defect at the hypersurface $u,y = 0$, $z^2 + w^2 = \lambda^{-1}$, i.e.\! a lightlike, circular string on a meridian of the cosmological horizon of de Sitter space\footnote{For an observer on the worldline $y,z,w = 0$, $x = x(t) > \lambda^{-1}$ the future/past horizon is the sphere $y^2 + z^2 + w^2 = \lambda^{-1}$, $x \pm t = 0$.}. Such a result is qualitatively different from 2+1 dimensions, where there is a pair of massless particles (similarly to the massive solution \cite{Jackiw:1984}) on the opposite points of the horizon \cite{Cai:1999}. However, it can simply be seen as the dimensional reduction of a circle to the degenerate case of a pair of points. 

The string can be better visualized in a different coordinate system, defined in analogy with \cite{Podolsky:1997}. To this end we first make the following transformation of coordinates
\begin{align}\label{eq:2.15}
t = \frac{1}{2\eta} \left(\lambda^{-1} - \eta^2 + (X + \lambda^{-\frac{1}{2}})^2 + Y^2 + Z^2\right)\,, \qquad x = \frac{1}{\sqrt{\lambda} \eta} (X + \lambda^{-\frac{1}{2}})\,, \nonumber\\ 
y = \frac{1}{\sqrt{\lambda} \eta} Y\,, \qquad z = \frac{1}{\sqrt{\lambda} \eta} Z\,, \qquad w = \frac{1}{2\eta} \left(\lambda^{-1} + \eta^2 - (X + \lambda^{-\frac{1}{2}})^2 - Y^2 - Z^2\right)
\end{align}
and then set $X = \rho \cos\phi \sin\theta$, $Y = \rho \sin\phi \sin\theta$, $Z = \rho \cos\theta$, so that the metric (\ref{eq:2.14}) takes the form
\begin{align}\label{eq:2.16}
ds^2 &= \frac{1}{\lambda \eta^2} \left(-d\eta^2 + d\rho^2 + \rho^2 \left(d\theta^2 + \sin^2\theta d\phi^2\right) \right) \nonumber\\ 
&- \frac{\varrho}{\sqrt{\lambda}} |\sin\phi \sin\theta| \left(\delta(\eta - \rho) (d\eta - d\rho)^2 + \delta(\eta + \rho) (d\eta + d\rho)^2\right)\,.
\end{align}
As one can observe, this metric describes a lightlike circular string lying at the great circle $\phi = 0, \pi$, $\theta \in [0, \pi]$ on the cosmological horizon $\rho = |\eta|$ of de Sitter space, whose metric is in the first line. The first term containing the Dirac delta is responsible for the string at times $\eta \geq 0$ and the other at times $\eta \leq 0$. 

The form of (\ref{eq:2.14}) is exactly the same as the metric (\ref{eq:2.7}) of a lightlike cosmic brane in 4+1-dimensional Minkowski space, as could be expected from the analogous situation for (\ref{eq:2.9}). This result easily generalizes to an arbitrary number of dimensions and thus we might say that a conical defect in the $n\!+\!1$-dimensional de Sitter space is a ``projection" of the corresponding defect in the embedding $(n\!+\!1)\!+\!1$-dimensional Minkowski space. The projection is understood in simple geometric terms: the $n\!-\!1$-dimensional hyperplane of a Minkowski defect is cutting through the $n$-dimensional sphere of a de Sitter spatial slice, determining the $n\!-\!2$-dimensional sphere of the projected defect.

\section{Summary}
In this work we have provided an exploration of the relation between the holonomies of conical defects in more than three spacetime dimensions and group valued momenta, which appear in scenarios of deformed relativistic symmetries. Our motivation was the well established fact that momentum space of point particles in 2+1-gravity, treated as conical defects, is a Lie group. We started by recalling how the motion of massive conical defects can be characterized by their holonomies, belonging to the Lorentz group. Generalizing the discussion to massless defects, we observed that holonomies describing such defects in $n\!+\!1$-dimensional Minkowski space, given by null rotations, can be seen as elements of a subgroup of the ${\rm AN}(n-1)$ group, which is the momentum space associated with $n$-dimensional $\kappa$-Minkowski space. In particular, for a fixed direction of propagation the space of holonomies of a massless defect is the above subgroup. This provides a partial ``physical" rendition of the group valued momenta emerging in the context of deformed relativistic symmetries, which were originally introduced as rather formal mathematical structures. At the same time, group momentum space (or, more broadly, momentum space with a nontrivial geometry) is a crucial ingredient of doubly special relativity and relative locality approaches to the problem of quantum gravity. Both these frameworks are based on heuristic assumptions and often use 2+1-gravity as a toy-model for their concepts. Thus our analysis of massless defects offers a certain practical setting to explore such ideas in a higher number of dimensions and in the context of the well studied $\kappa$-Minkowski space. Furthermore, similar group momentum spaces are very likely associated with defects coupled to the BF-theory \cite{Baez:2007}, which is a topological field analogue of gravity in more than three dimensions. In this paper we also discussed the case of massless conical defects in 3+1-dimensional de Sitter space, showing how they can be seen as closed lightlike strings and how their metric is strictly related to the one of massless defects in 4+1-dimensional Minkowski space. This led us to conjecture that the same holonomies, belonging to the ${\rm AN}(3)$ group, which characterize the latter could completely capture the motion of massless defects in de Sitter space and that such a statement could be generalized to more spacetime dimensions. A proof of this conjecture is postponed to future studies.

\section*{Acknowledgments}
We would like to thank M. van de Meent for very useful correspondence and J. Kowalski-Glikman for comments on the manuscript. The work of MA is supported by a Marie Curie Career Integration Grant within the 7th European Community Framework Programme and in part by a grant from the John Templeton Foundation. TT acknowledges the support by the Foundation for Polish Science International PhD Projects Programme co-financed by the EU European Regional Development Fund and the additional funds provided by the National Science Center under the agreements no. DEC-2011/02/A/ST2/00294 and 2014/13/B/ST2/04043.

\appendix
\section{Representations of $\kappa$-Minkowski generators in 3+1 dimensions}
For the purpose of illustration we write here down the three independent $4 \times 4$ matrix representations of the $\mathfrak{an}(2)$ generators that can be used to reconstruct holonomies of massless defects for our choice of vectors ${\bf n}_{(a)}$ and ${\bf d}_{(b)}$, $a,b = x,y,z$ which we introduced in Section~III. Defects with the vectors ${\bf n}_{(x)}$ and ${\bf d}_{(1)} \equiv {\bf d}_{(z)}$, ${\bf d}_{(2)} \equiv {\bf d}_{(y)}$ correspond to the representation $X^{(x)}_0 = K_x$, $X^{(x)}_1 = K_y - J_z$, $X^{(x)}_2 = K_z + J_y$, where
\begin{align}\label{eq:A.1}
X^{(x)}_0 = -i\left(
\begin{array}{cccc}
0\! & 1\! & 0\! & 0 \\ 
1\! & 0\! & 0\! & 0 \\ 
0\! & 0\! & 0\! & 0 \\ 
0\! & 0\! & 0\! & 0 
\end{array}
\right)\,, \qquad 
X^{(x)}_1 = i\left(
\begin{array}{cccc}
0\! & 0\! & -1\! & 0 \\ 
0\! & 0\! & -1\! & 0 \\ 
-1\! & 1\! & 0\! & 0 \\ 
0\! & 0\! & 0\! & 0 
\end{array}
\right)\,, \qquad  
X^{(x)}_2 = i\left(
\begin{array}{cccc}
0\! & 0\! & 0\! & -1 \\ 
0\! & 0\! & 0\! & -1 \\ 
0\! & 0\! & 0\! & 0 \\ 
-1\! & 1\! & 0\! & 0 
\end{array}
\right)\,;
\end{align}
the vectors ${\bf n}_{(y)}$ and ${\bf d}_{(1)} \equiv {\bf d}_{(z)}$, ${\bf d}_{(2)} \equiv {\bf d}_{(x)}$ correspond to $X^{(y)}_0 = K_y$, $X^{(y)}_1 = K_x + J_z$, $X^{(y)}_2 = K_z - J_x$, where
\begin{align}\label{eq:A.2}
X^{(y)}_0 = -i\left(
\begin{array}{cccc}
0\! & 0\! & 1\! & 0 \\ 
0\! & 0\! & 0\! & 0 \\ 
1\! & 0\! & 0\! & 0 \\ 
0\! & 0\! & 0\! & 0 
\end{array}
\right)\,, \qquad 
X^{(y)}_1 = i\left(
\begin{array}{cccc}
0\! & -1\! & 0\! & 0 \\ 
-1\! & 0\! & 1\! & 0 \\ 
0\! & -1\! & 0\! & 0 \\ 
0\! & 0\! & 0\! & 0 
\end{array}
\right)\,, \qquad  
X^{(y)}_2 = i\left(
\begin{array}{cccc}
0\! & 0\! & 0\! & -1 \\ 
0\! & 0\! & 0\! & 0 \\ 
0\! & 0\! & 0\! & -1 \\ 
-1\! & 0\! & 1\! & 0 
\end{array}
\right)\,;
\end{align}
the vectors ${\bf n}_{(z)}$ and ${\bf d}_{(1)} \equiv {\bf d}_{(y)}$, ${\bf d}_{(2)} \equiv {\bf d}_{(x)}$ correspond to $X^{(z)}_0 = K_z$, $X^{(z)}_1 = K_x - J_y$, $X^{(z)}_2 = K_y + J_x$, where
\begin{align}\label{eq:A.3}
X^{(z)}_0 = K_z = -i\left(
\begin{array}{cccc}
0\! & 0\! & 0\! & 1 \\ 
0\! & 0\! & 0\! & 0 \\ 
0\! & 0\! & 0\! & 0 \\ 
1\! & 0\! & 0\! & 0 
\end{array}
\right)\,, \qquad 
X^{(z)}_1 = i\left(
\begin{array}{cccc}
0\! & -1\! & 0\! & 0 \\ 
-1\! & 0\! & 0\! & 1 \\ 
0\! & 0\! & 0\! & 0 \\ 
0\! & -1\! & 0\! & 0 
\end{array}
\right)\,, \qquad 
X^{(z)}_2 = i\left(
\begin{array}{cccc}
0\! & 0\! & -1\! & 0 \\ 
0\! & 0\! & 0\! & 0 \\ 
-1\! & 0\! & 0\! & 1 \\ 
0\! & 0\! & -1\! & 0 
\end{array}
\right)\,.
\end{align}
$J_a$, $K_a$, $a = x,y,z$ respectively denote the generators of rotations and boosts.

\section{Conical defects in anti-de Sitter space}
For completeness let us also discuss conical defects in anti-de Sitter space, with $\Lambda < 0$. In this case the situation is very similar to the de Sitter case and therefore we will be very brief. We again study the generalization of a massless defect from 2+1-dimensions \cite{Cai:1999} to 3+1 dimensions. 

For brevity we set $\Lambda = -3$. Then the metric of a static massive defect in 3+1 dimensions \cite{Dehghani:2002,Mello:2012}, analogous to a particle in 2+1 dimensions \cite{Jackiw:1984}, in static anti-de Sitter coordinates has the form
\begin{align}\label{eq:A.4}
ds^2 = -(1 + r^2) d\tau^2 + (1 + r^2)^{-1} dr^2 + r^2 \left(d\theta^2 + (1 - 4\mu)^2 \sin^2\theta d\phi^2\right)
\end{align}
and describes a cosmic string of rest energy density $\mu$ and the deficit angle $8\pi\mu$. One may expand it to the first power in $\mu$ and transform to embedding coordinates
\begin{align}\label{eq:A.5}
t = \sqrt{1 + r^2} \sin\tau\,, \qquad w = \pm\sqrt{1 + r^2} \cos\tau\,, \nonumber\\ 
x = r \cos\phi \sin\theta\,, \qquad y = r \sin\phi \sin\theta\,, \qquad z = r \cos\theta\,,
\end{align}
which satisfy the $(2,3)$-hyperboloid condition $-1 = -t^2 + x^2 + y^2 + z^2 - w^2$. We subsequently perform the Aichelburg-Sexl boost, obtaining eventually (there is no difference for $\Lambda \neq -3$)
\begin{align}\label{eq:A.6}
ds^2 = 2 du dv + dy^2 + dz^2 - dw^2 - \sqrt{2}\varrho |y| \delta(u) du^2\,.
\end{align}
The above metric corresponds to a curvature singularity at the hypersurface $u,y = 0$, $-z^2 + w^2 = 1$, i.e.\! a lightlike hyperbolic string. The hyperbola has two branches but these actually belong to the same worldsheet, analogously to 2+1 dimensions \cite{Cai:1999} (in contrast to the de Sitter case), and therefore it is a single defect. Moreover, the form of (\ref{eq:A.6}) is different from the metric (\ref{eq:2.7}) of a massless defect in 4+1-dimensional Minkowski space. For an additional insight we change the coordinates to
\begin{align}\label{eq:A.7}
t = \frac{1}{2Z} \left(1 - \eta^2 + (X + 1)^2 + Y^2 + Z^2\right)\,, \qquad x = \frac{1}{Z} (X + 1)\,, \nonumber\\ 
y = \frac{1}{2Z} \left(1 + \eta^2 - (X + 1)^2 - Y^2 - Z^2\right)\,, \qquad z = \frac{1}{Z} Y\,, \qquad w = \frac{1}{\eta} Z
\end{align}
and once more to $X = \rho \cos\phi \sin\theta$, $Y = \rho \sin\phi \sin\theta$, $Z = \rho \cos\theta$. Restoring $\Lambda \neq -3$, we find that
\begin{align}\label{eq:A.8}
ds^2 &= \frac{3}{|\Lambda| \rho^2 \cos^2\theta} \left(-d\eta^2 + d\rho^2 + \rho^2 \left(d\theta^2 + \sin^2\theta d\phi^2\right) \right) \nonumber\\ 
&- \varrho \sqrt{\frac{3}{|\Lambda|}} \frac{|\sin\phi \sin\theta|}{\cos^2\theta} \left(\delta(\eta - \rho) (d\eta - d\rho)^2 + \delta(\eta + \rho) (d\eta + d\rho)^2\right)\,.
\end{align}
The string lies on the surface $\rho = |\eta|$, which is a hyperboloid in anti-de Sitter space. Again, one term with the Dirac delta is responsible for the defect at times $\eta \geq 0$ and the other at times $\eta \leq 0$.

\end{document}